# How to Recognize Clustering of Luminescent Defects in Single-Wall Carbon Nanotubes

Finn L. Sebastian[1], Simon Settele[1], Han Li[2,3], Benjamin S. Flavel[4], and Jana Zaumseil[1,*]

[1]Institute for Physical Chemistry, Universität Heidelberg, D-69120 Heidelberg, Germany

[2]Department of Mechanical and Materials Engineering, University of Turku, FI-20014 Turku, Finland

[3]Turku Collegium for Science, Medicine and Technology, University of Turku, FI-20520 Turku, Finland

[4]Institute of Nanotechnology, Karlsruhe Institute of Technology, D-76131 Karlsruhe, Germany

**Corresponding Author**

*E-mail: zaumseil@uni-heidelberg.de

**Abstract**

Semiconducting single-wall carbon nanotubes (SWCNTs) are a promising material platform for near-infrared in-vivo imaging, optical sensing, and single-photon emission at telecommunication wavelengths. The functionalization of SWCNTs with luminescent defects can lead to significantly enhanced photoluminescence (PL) properties due to efficient trapping of highly mobile excitons and red-shifted emission from these trap states. Among the most studied luminescent defect types are oxygen and aryl defects that have largely similar optical properties. So far, no direct comparison between SWCNTs functionalized with oxygen and aryl defects under identical conditions has been performed. Here, we employ a combination of spectroscopic techniques to quantify the number of defects, their distribution along the nanotubes and thus their exciton trapping efficiencies. The different slopes of Raman D/G$^+$ ratios versus calculated defect densities from PL quantum yield measurements indicate substantial dissimilarities between oxygen and aryl defects. Supported by statistical analysis of single-nanotube PL spectra at cryogenic temperatures it reveals clustering of oxygen defects. The clustering of 2-3 oxygen defects, which act as a single exciton trap, occurs irrespective of the functionalization method and thus enables the use of simple equations to determine the density of oxygen defects and oxygen defect clusters in SWCNTs based on standard Raman spectroscopy. The presented analytical approach is a versatile and sensitive tool to study defect distribution and clustering in SWCNTs and can be applied to any new functionalization method.

**Introduction**

The covalent functionalization of single-wall carbon nanotubes (SWCNTs) with luminescent defects has emerged as a promising strategy to improve and tune their near-infrared luminescence properties for optical sensing,[1-4] high contrast in-vivo imaging in the second biological window,[5,6] high-purity single-photon emission,[7-9] and optoelectronic devices operating at telecommunication wavelengths.[10-12] Regardless of their exact chemical composition or binding configuration, luminescent defects act as zero-dimensional trap sites for mobile $E_{11}$ excitons diffusing along the SWCNT lattice.[13,14] Photoluminescence (PL) from these energetically lower states occurs at wavelengths even further red-shifted into the near-infrared (nIR) than $E_{11}$ and with fluorescence lifetimes of hundreds of picoseconds. The energy difference between the mobile $E_{11}$ exciton and defect emission (i.e., optical trap depth, $\Delta E_{opt}$) ranges from 100 to 300 meV. Luminescent defects are able to trap and localize excitons at room temperature and thus prevent nonradiative decay at quenching sites (e.g., nanotube ends). The suppression of this non-radiative decay path considerably enhances the photoluminescence quantum yields (PLQYs) of SWCNTs within a narrow window of optimal defect density.[15-18] Most commonly these defects are created, characterized and theoretically investigated in chiral (6,5) carbon nanotubes due to their abundance, easy purification and high reactivity.[19-21] Hence, we will focus on (6,5) SWCNTs from here on.

Among the various types of luminescent defects in (6,5) SWCNTs, oxygen and aryl defects have been the main focus of experimental and theoretical studies, as they are easily introduced using reactive oxygen species (ROS)[6,20,22,23] or aryldiazonium salts (Figure 1a).[14,24-26] In both cases, the emission wavelength of the defect PL is mainly determined by the binding configuration of the introduced defect moieties on the SWCNT lattice. For oxygen defects, three thermodynamically stable binding configurations can be distinguished.[13] In one of them, the oxygen atom forms an epoxide, while the other two are ether-type configurations with different optical trap depths depending on the carbon-carbon bonds in either circumferential (ether-D) or longitudinal (ether-L) orientation with respect to the SWCNT axis, as shown schematically in Figure 1b. Similar to that, two stable and experimentally observed configurations of sp³ defects with distinct PL emission wavelengths are created upon binding of aryl groups to the nanotube lattice and formation of two sp³ carbons in different ortho (e.g., aryl-L$_{90}$ see Figure 1b) or para positions.[27-29]

Luminescent oxygen and aryl defects exhibit similar optical properties in addition to their equally red-shifted emission, such as nonlinear scaling of the defect emission intensity with excitation power[30] and a significant energy offset between the optical and thermal trap depths.[31] However, large deviations were found for the brightening factors for SWCNTs functionalized with oxygen or aryl sp³ defects. The reported absolute values vary, but, typically oxygen defects lead to an increase in total PLQY by a factor of 2-3,[6, 20] whereas aryl defects can cause a PLQY increase by a factor of >5, depending on the functional group and initial SWCNT length.[15, 17, 24] A reversed order of bright and dark exciton states for oxygen and aryl defects was considered as a possible explanation.[14, 31] Additionally, the introduction of exciton quenching sites due to unintentional over-oxidation of SWCNTs under harsh conditions during functionalization with ROS may play a role.[6, 20] The efficient capture of mobile $E_{11}$ excitons followed by defect emission depends on the distribution and trap depth of different defects on the nanotubes lattice, which is usually assumed to be uniform. However, clustering of defects may occur depending on the functionalization method and should have a substantial impact on emission properties relevant for applications in sensing, imaging and optoelectronics.

To this end, we present a comparative study of oxygen- and aryl-functionalized (6,5) SWCNTs and demonstrate how to recognize clustering of luminescent defects. We employ a combined analytical approach of Raman spectroscopy and spectrally-resolved PLQY measurements to quantify the absolute density of luminescent exciton quenching sites on the carbon nanotube lattice for different functionalization reactions. This method provides precise defect densities (per µm of nanotube) for aryl sp³ defects in different SWCNT species and independent of the Raman excitation laser wavelength as shown previously.[32, 33] However, we find a striking difference in exciton trapping efficiency for SWCNTs with luminescent oxygen defects at identical levels of functionalization to aryl defects as corroborated by a statistical analysis of single-nanotube PL spectroscopy at cryogenic temperature. We attribute these findings to clustering of oxygen defects at length scales similar to or smaller than the mobile exciton size, whereas aryl defects act mostly as separate exciton traps. We propose the presented approach as a robust and powerful tool to identify and quantify inhomogeneities in defect distribution and defect clustering in functionalized SWCNTs.

**Results and Discussion**

To quantify and directly compare different luminescent oxygen and aryl defects, purified dispersions of (6,5) SWCNTs stabilized in water by sodium dodecyl sulfate (SDS) were obtained through aqueous two-phase extraction as reported previously.[32] The nearly monochiral (6,5) SWCNT dispersions were characterized by UV-Vis-nIR absorption and PL excitation-emission spectroscopy (Figures S1 and S2, ESI†). The functionalization of (6,5) SWCNTs with luminescent defects was performed using a variety of synthetic methods (see Figure 1a). Oxygen defects were introduced via three previously reported procedures: ozonation of SWCNTs and subsequent photoconversion with visible light[20], UV-light driven dissociation of sodium hypochlorite (NaOCl)[6], and with a Fenton-like reaction using copper(II) sulfate ($CuSO_4$) and sodium ascorbate (NaAsc)[23] (for details, see Experimental Methods, ESI†). While oxygen functionalization was shown to initially introduce ether-L, ether-D and epoxide-L defects with different trap depths (Figure 1b),[13] the epoxide defects mostly rearrange to the thermodynamically favored ether-D configuration upon exposure of SWCNTs to light.[23] The covalent functionalization with aryl defects was accomplished using 4-nitrobenzenediazonium tetrafluoroborate ($DzNO_2$), which forms radical species that attack the $sp^2$-hybridized nanotube lattice leading to $sp^3$ carbons with a predominantly aryl-$L_{90}$ configuration.[28]

The degree of functionalization was tuned by varying the concentration of the respective reactants. In the case of SWCNT ozonation, the characteristic UV absorption peak of ozone at 260 nm was monitored to adjust its concentration in water prior to addition to the nanotube dispersion, as shown in Figure S3, ESI†. Importantly, all functionalization reactions were performed on the same batch of SWCNTs and in the same surfactant (SDS) to ensure comparability of the subsequent characterization. The successful introduction of luminescent defects was confirmed by the emergence of defect-induced emission peaks at ~1120 nm (ether-D defect) and ~1140 nm (aryl-$L_{90}$ defect) as shown in Figures 1c-f (for absolute PL spectra, see Figure S4, ESI†). These defect emission peaks are commonly labelled based on their spectral position with respect to the $E_{11}$ emission and not their precise molecular origin. Hence, for both oxygen and aryl defects we refer to the main defect emission peaks as $E_{11}*$. For all functionalization methods, a linear increase of the $E_{11}*/E_{11}$ PL intensity ratio with the reactant concentration was observed (Figure S5, ESI†). Note that the Fenton-like reaction (Figure 1c) leads to a narrower $E_{11}*$ emission peak, which we attribute to the very mild reaction conditions with fewer possible side-reactions.

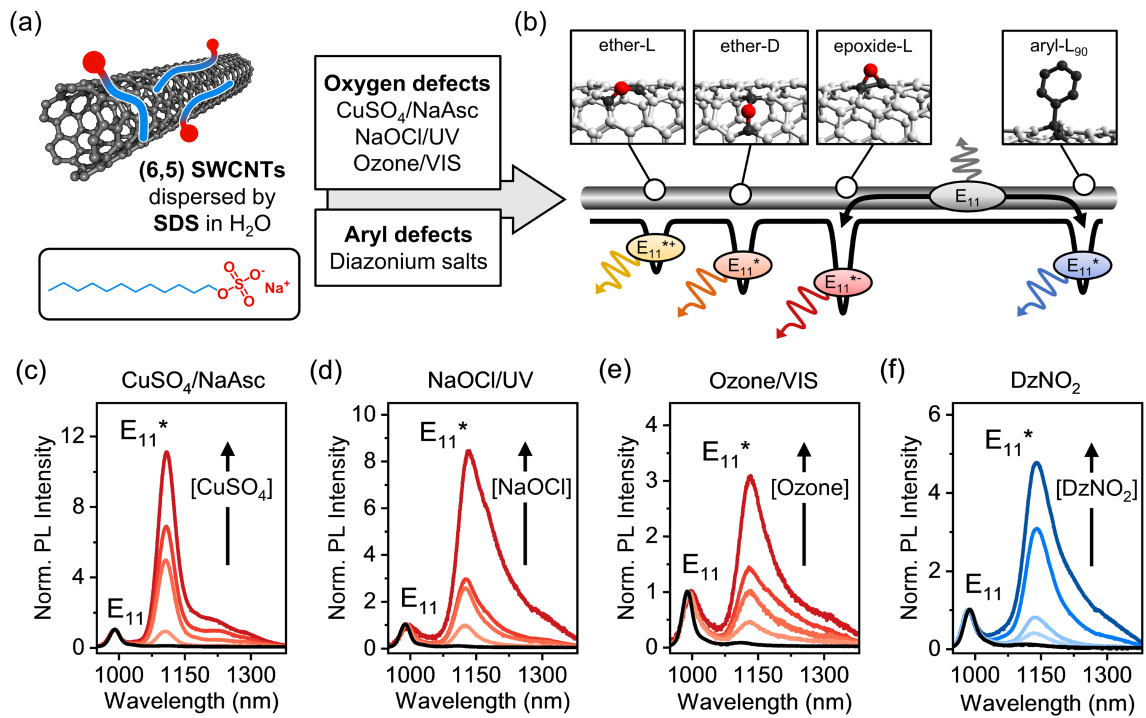

**Figure 1.** **(a)** Covalent functionalization of (6,5) SWCNTs with luminescent defects in aqueous dispersion with surfactant SDS. Oxygen defects are created using $CuSO_4$/NaAsc, NaOCl/UV light, or ozonation followed by SWCNT excitation with visible light. Treatment of SWCNTs with 4-nitrobenzenediazonium tetrafluoroborate ($DzNO_2$) introduces aryl $sp^3$ defects. **(b)** Schematic of different types of oxygen (ether-L, ether-D, and epoxide-L) and $sp^3$ defects (aryl-$L_{90}$) that act as trapping sites with different depths for mobile $E_{11}$ excitons. **(c-f)** Normalized (to $E_{11}$) PL spectra of pristine (black) and functionalized (6,5) SWCNTs in aqueous dispersion.

After functionalization, a surfactant exchange of the SWCNTs to 1% (w/v) sodium deoxycholate (DOC) was carried out via spin-filtration to remove reaction by-products and generate stable dispersions for further analysis. Drop-cast films of all dispersions were analyzed by resonant Raman spectroscopy (laser excitation wavelength $\lambda_{exc}$ = 532 nm). Figures 2a-d show averaged Raman spectra for all functionalization methods at different defect densities (corresponding to samples in Figure 1c-f). The Raman D-mode intensity is proportional to the number of point defects in the SWCNT lattice, whereas the intensity of the $G^+$-mode originates from the longitudinal optical phonon of the $sp^2$-hybridized nanotube lattice.[34] Thus, the Raman $D/G^+$ ratio increases with functionalization and can be used as a relative metric to estimate the density of luminescent defects. For different sp³ aryl defects in (6,5) SWCNTs - but not yet for oxygen defects - a quantitative correlation was established, vide infra.[32, 33]

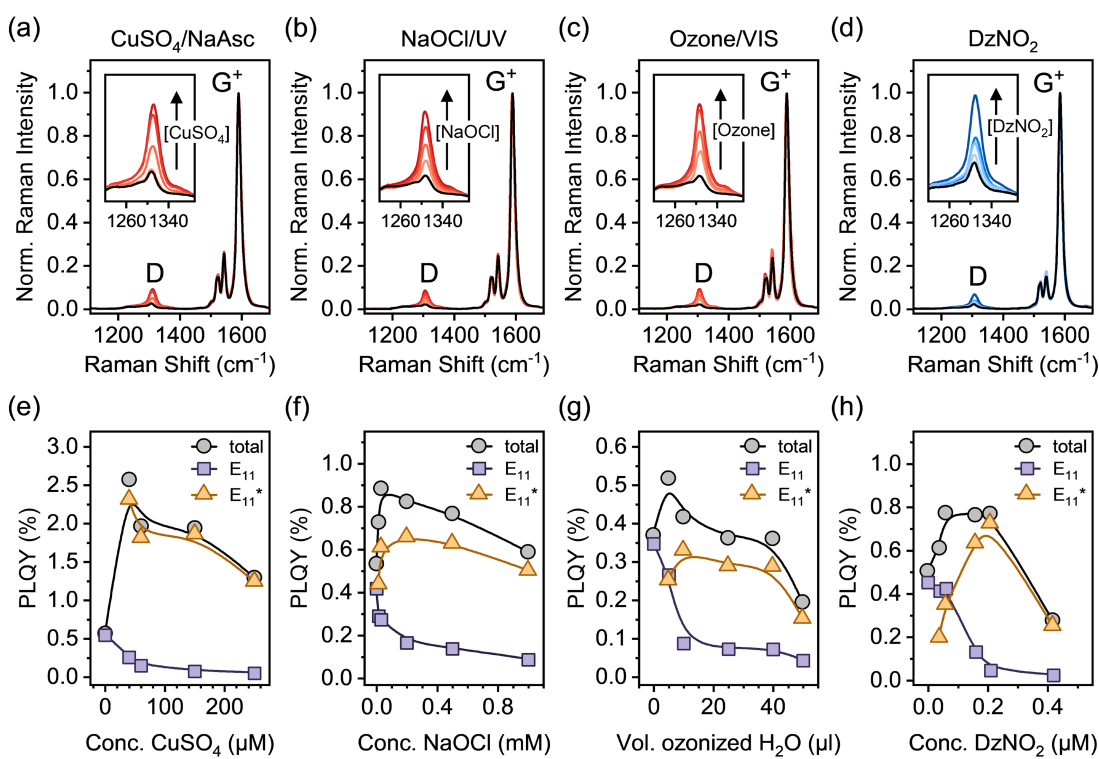

**Figure 2**. **(a-d)** Normalized (to G$^+$ peak) averaged Raman spectra of pristine (black) and different functionalized (6,5) SWCNTs ($\lambda_{exc}$ = 532 nm, >3600 individual spectra for each sample). The insets show the evolution of the defect-related D-mode with increasing degree of functionalization. **(e-f)** Corresponding total and spectrally resolved PLQYs for pristine and different functionalized (6,5) SWCNTs in aqueous dispersion depending on reactant concentration (after surfactant exchange to 1 % (w/v) DOC).

Treatment of nanotubes with oxidizing agents such as ozone, hypochlorite or CuSO$_4$/NaAsc, although at low concentrations, can lead to slight p-doping, which is also reflected in their Raman spectra and may complicate the analysis for defect quantification. Signatures of doping, i.e., increased intensities of the E$_2$ mode as well as broadening of the G$^+$- and 2D-mode linewidth were observed for some oxygen-functionalized SWCNTs.[35, 36] To separate these effects, films of pristine and functionalized (6,5) SWCNTs were intentionally p-doped by immersion in toluene solutions of the molecular oxidant F$_4$TCNQ (for details, see Experimental Methods, ESI†).[37, 38] While the typical signatures of p-doping are observed (Figure S6, ESI†), no changes of the integrated Raman D/G$^+$ ratio occur (Figure S7, ESI†), indicating that low-level p-doping does not interfere with the defect quantification for oxygen-functionalized SWCNTs by Raman spectroscopy.

Complementary to the relative defect densities obtained by Raman spectroscopy, absolute defect densities can be calculated within the framework of the diffusion-limited contact quenching (DLCQ) model. Within this model, exciton diffusion in SWCNTs is assumed to be governed exclusively by radiative decay of highly mobile $E_{11}$ excitons or non-radiative decay at quenching sites, e.g. nanotube ends.[17, 39] Covalent functionalization with luminescent defects introduces new loss channels for $E_{11}$ excitons, leading to an additional decrease of the $E_{11}$ PLQY. This enables the calculation of luminescent defect densities, provided that PLQYs can be determined reliably. The combination of relative Raman $D/G^+$ ratios and defect densities calculated from the spectrally resolved $E_{11}$ PLQY of pristine and functionalized SWCNTs is a powerful tool to compare and investigate different types of luminescent defects as both techniques are highly sensitive to small changes in the electronic structure of nanotubes. For $sp^3$ aryl defects created by different synthetic routes this method was already shown to provide a robust correlation for defects densities up to ~ 40 $\mu m^{-1}$ and thus enables facile quantification of the defect density by Raman spectroscopy alone.[32, 33]

To compare and expand this correlation to oxygen defects, we determined the absolute PLQYs of (6,5) SWCNTs functionalized with oxygen and aryl defects and calculated the defect densities from the measured decrease in $E_{11}$ PLQY (for details on PLQY measurements and defect density calculation see Experimental Methods, ESI†). Spectrally resolved PLQYs for pristine and functionalized SWCNTs are presented in Figure 2e-h. At low degrees of functionalization, an increase of the total PLQY is observed irrespective of the functionalization method. Efficient exciton trapping and radiative relaxation via $E_{11}*$ emission overcompensates for the simultaneous decrease in $E_{11}$ emission. At medium to high levels of functionalization, the total PLQY decreases again, as the defects start to significantly disturb the electronic structure of the pristine SWCNT lattice.

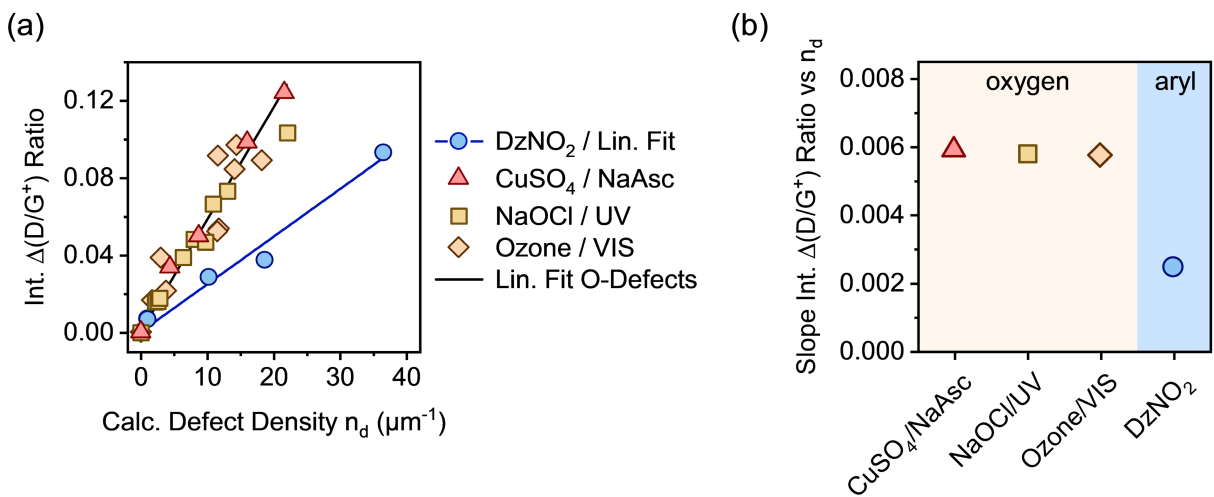

**Figure 3**. **(a)** Integrated Raman Δ(D/G$^+$) ratios versus calculated defect densities ($n_d$) based on PLQY for (6,5) SWCNTs functionalized by different methods creating oxygen or aryl sp$^3$ defects. Linear fits for oxygen defects, combined ($R^2 = 0.97$) and aryl sp$^3$ defects ($R^2 = 0.98$) **(b)** Extracted slopes of the integrated Δ(D/G$^+$) ratio versus calculated defect density for each functionalization method.

In analogy to our previous studies on aryl defects,[32, 33] the change of the integrated Raman D/G$^+$ ratios (Δ(D/G$^+$)) was correlated with the calculated defect densities $n_d$ from $E_{11}$ PLQY measurements (see Figure 3a). Using the Δ(D/G$^+$) ratio instead of the absolute values generally ensures higher comparability of the Raman data across different SWCNT batches and starting materials. In this work, the same batch of nanotube material was used for all samples. A linear dependence of the Δ(D/G$^+$) ratio on the calculated density of luminescent oxygen defects was obtained for all samples. For luminescent aryl defects, the slope was the same as previously reported for (6,5) SWCNTs.[32, 33] In contrast to that, a strong deviation from this gradient is evident for all oxygen-functionalized nanotubes. Regardless of the functionalization method, all samples of (6,5) SWCNTs with luminescent oxygen defects exhibit a slope that is larger by a factor of ~2.3 than that for the aryl defects (see Figure 3b; for individual linear fits see Figure S8, ESI†). Similar differences are also observed for other defect-related Raman peaks, such as the intermediate frequency modes (IFMs)[33, 40] presented in Figures S9 - S11, ESI†.

Clearly, despite very similar spectral signatures in PL and Raman spectroscopy, there must be a fundamental difference between luminescent oxygen and aryl defects in SWCNTs and their interaction with mobile excitons. Importantly, the calculation of defect densities based on the DLCQ model only considers the diffusion constant and radiative lifetime of the $E_{11}$ exciton but no defect-specific photophysical processes.[17, 41] Hence, it gives direct access to the $E_{11}$ exciton quenching probability of local defects on the SWCNT lattice. However, the observed deviation of the slopes of the Raman $D/G^+$ ratios indicate either a different D-mode Raman cross-section or higher defect densities for oxygen-functionalized compared to aryl-functionalized nanotubes.

Previous work on defective carbon-based materials indicated that the intensity of the disorder-related Raman D-mode is not particularly sensitive to the actual structure (functional group or hybridization) of point-like lattice defects.[34, 42] As all types of oxygen and aryl defects involve binding to exactly two carbon atoms, the defect size and thus the probability of second-order scattering processes (i.e., Raman D-mode cross-section) should be equally insensitive to the precise chemical nature of the point-like defects. This assumption only leaves a higher number of structural defects than calculated from PLQY measurements as a possible explanation, which could be caused by clustering of the oxygen defects. The possibility of clustering of defects during covalent functionalization of carbon nanotubes was considered before,[43] but only shown to occur in additional alkylation reactions on already functionalized SWCNTs.[44] Furthermore, theoretical studies on SWCNT shortening by chemical oxidation with hydrogen peroxide suggest that ROS preferentially attack the carbon nanotube lattice at already defective areas, which may promote clustering.[45]

For a cluster of defects with an average spacing similar to or smaller than the size of the exciton (i.e., few nanometers[17, 46]), several defects may act as a single quenching site for the mobile excitons in the DLCQ model. Furthermore, collective exciton states of coupled defects were proposed for closely spaced luminescent defects.[47-50] Irrespective of the actual local electronic structure, the exciton trapping efficiency probably does not simply scale linearly with the absolute number of individual luminescent defects. Their precise distribution along the SWCNT lattice will strongly influence the correlation between lattice defects as determined from the Raman $D/G^+$ ratio and the calculated defect density from $E_{11}$ PLQY measurements.

To directly assess the possibility of defect clustering for oxygen defects, we performed PL spectroscopy on individual oxygen-functionalized SWCNTs embedded in a polymer matrix at cryogenic temperatures (4.7 K). Under these conditions SWCNTs show strong localization of even the $E_{11}$ excitons due to slight variations in the local dielectric environment.[51,52] Luminescent defects are similarly influenced by dielectric fluctuations of the environment. Thus, the defect emission energies even of luminescent defects with identical binding configuration vary slightly. These differences can be resolved at cryogenic temperatures due to the strongly reduced linewidths. Hence, it is possible to quantify the number of luminescent defects on a single nanotube by simply counting the different peaks, as demonstrated previously for aryl defects in polymer-wrapped SWCNTs.[32] The statistical analysis of the number of defect emission peaks provides direct access to the average number of individual emissive sites per nanotube. Together with atomic force microscopy (AFM) to determine the average nanotube length, the density of luminescent defects per µm can be obtained and compared to the results obtained from PLQY measurements and the DLCQ model.

To perform this analysis, oxygen-functionalized (6,5) SWCNTs were transferred to toluene solutions of the fluorene-bipyridine copolymer PFO-BPy, which exhibits a high dispersion efficiency for (6,5) SWCNTs.[21] After transfer of highly dilute nanotube dispersions into a polystyrene matrix, single-nanotube PL spectra were acquired at 4.7 K in an optical cryostat (for details see Experimental Methods, ESI†). Representative PL spectra for individual (6,5) SWCNTs with $E_{11}^*$ oxygen defects are displayed in Figures 4a-d. Single defects can be identified as narrow peaks in the respective spectral region characteristic for a certain binding configuration (1030-1100 nm for $E_{11}^{*+}$, 1100-1200 nm for $E_{11}^*$, 1200-1300 nm for $E_{11}^{*-}$). Figures S12 and S13 (ESI†) show additional single-SWCNT PL spectra.

About 30% of the measured SWCNTs exhibited defect PL emission in the $E_{11}^{*+}$ and $E_{11}^{*-}$ spectral region (Figures 5e-g). The $E_{11}^{*+}$ emission likely occurs from longitudinally oriented ether-type defects (ether-L) with a shallower exciton trap depth ($\Delta E_{opt}$ ~50 meV, compared to ~140 meV for $E_{11}^*$ and ~250 meV for $E_{11}^{*-}$).[13] This defect is also often referred to as the $Y_1$ band in PL spectra of SWCNTs and becomes more prominent in thin films of SWCNTs on substrates with polar surfaces (e.g., $SiO_2$ or glass) upon annealing.[53] Emission from this shallow exciton trap state is low at room temperature but becomes significant under cryogenic conditions. It is plausible that thermal detrapping of excitons from these shallow traps back to the $E_{11}$ level is dominant at room temperature.[31] Hence, $E_{11}^{*+}$ defects should not contribute to the decrease in $E_{11}$ PLQY to the same degree as deeper exciton traps ($E_{11}^*$ and $E_{11}^{*-}$) while still leading to a higher $\Delta(D/G^+)$ ratio.

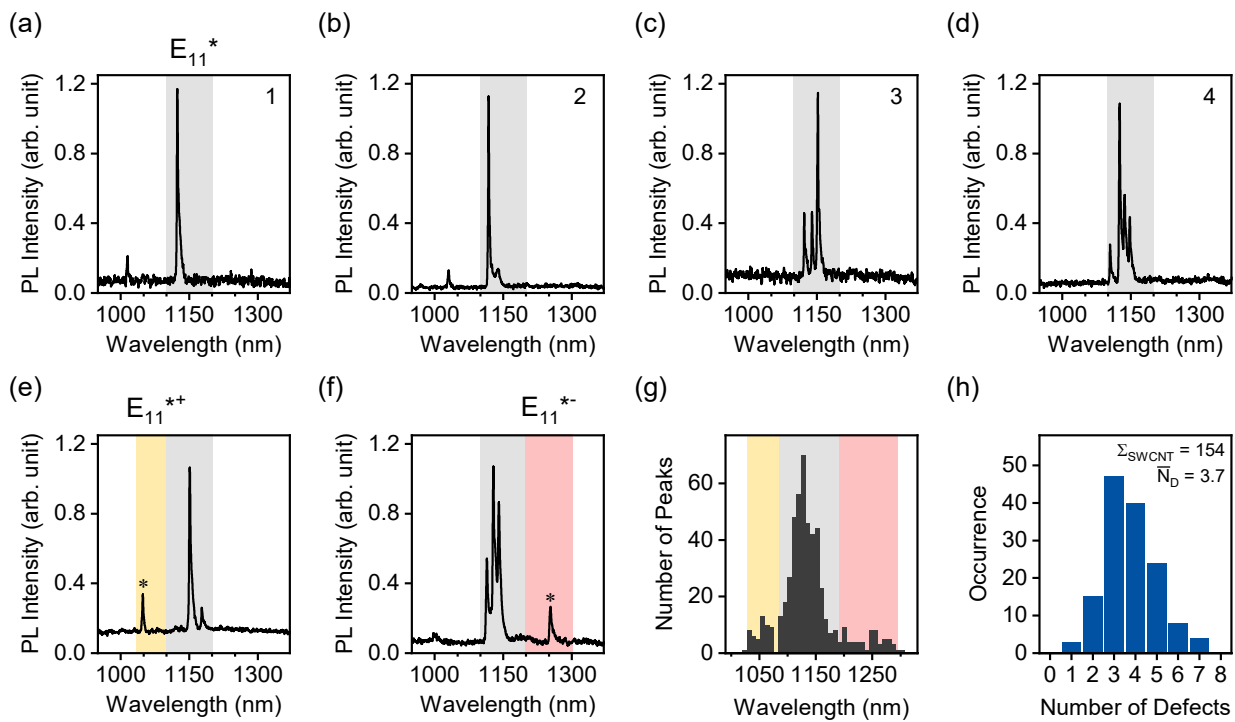

**Figure 4**. **(a-d)** Low-temperature (4.7 K) PL spectra of individual (6,5) SWCNTs functionalized with CuSO$_4$/NaAsc, embedded in a polystyrene matrix. The number of oxygen defect sites per nanotube (top right corner) was determined as the number of peaks in the spectral region of the E$_{11}$* emission (grey area). **(e, f)** Representative low-temperature spectra where PL emission from the E$_{11}$*$^+$ (yellow area) and E$_{11}$*$^-$ (pink area) defect configurations is observed. **(g)** Spectral distribution of all counted defect PL peaks. **(h)** Defect PL peak histogram of 154 individual (6,5) SWCNTs with an average defect density of 3.7 defects per nanotube.

A full statistical analysis of over 150 single-SWCNT PL spectra gave an average value of 3.7 luminescent defects per nanotube, with a minimum of one and a maximum of seven defects per nanotube, as shown in Figure 4h. AFM length statistics of the sample yielded an average nanotube length of 310 nm (compare Figure S14, ESI†). Combining these values gives an average defect density of 11.9 μm$^{-1}$ for oxygen functionalized SWCNTs, which is 2.4 times larger than the calculated defect density of 4.9 μm$^{-1}$ from the E$_{11}$ PLQY of the same sample. When the same counting method was applied to sp³ aryl defects in previous studies, the defect densities aligned remarkably well with the numbers extracted with the DLCQ model.[32] This ratio of real and expected defect density is also remarkably close to the factor of 2.3 by which the Δ(D/G$^+$) vs n$_d$ slope deviates for oxygen and aryl defects as shown in Figure 3a.

Clearly, the correlation between the Raman $\Delta(D/G^+)$ ratio and the defect density – determined from $E_{11}$ PLQYs or by counting the defects on individual SWCNTs – is strikingly different for oxygen and aryl sp$^3$ defects. As argued above, this deviation could be explained with the preferential formation of oxygen defect clusters by ROS-based functionalization methods. Assuming that 2 to 3 defects form a cluster smaller than the size of a mobile exciton, these excitons would interact with such clustered oxygen defects as one collective trap at room temperature. This would result in an apparently lower trapping efficiency per structural defect as measured by the Raman D/G$^+$ ratio. Thus, higher $E_{11}$ PLQYs and lower calculated defect densities would be observed for the same number of structural defects. Conversely, at cryogenic conditions even minor differences in the local dielectric environment of oxygen defect clusters due to inhomogeneities of the polymer matrix would distinguish these traps from each other and enable a direct quantification. At low temperatures, exciton detrapping is suppressed, and even individual defects with low exciton trapping probabilities within a larger cluster can be identified in single-nanotube PL spectra.

The remarkable agreement between data obtained independently by Raman/PLQY and cryo-PL/AFM measurements suggests that the quantification of the number of oxygen defect clusters can be performed by adapting the equation derived for aryl defects.[32] The number of collective exciton traps in the form of oxygen clusters is then

$$n_{\text{O-Cluster}}(\mu m^{-1}) = 172 \ \mu m^{-1} \cdot \Delta\left(\frac{I_D}{I_{G^+}}\right), \tag{1}$$

where $\Delta(I_D/I_{G^+})$ represents the change of the integrated Raman D/G$^+$ ratio of the functionalized (6,5) SWCNT sample compared to the pristine samples (for resonant excitation at 532 nm). Consequently, the actual number of individual oxygen atom defects can be calculated with

$$n_{\text{O-Defect}}(\mu m^{-1}) = 405 \ \mu m^{-1} \cdot \Delta\left(\frac{I_D}{I_{G^+}}\right), \tag{2}$$

which is, within the margin of experimental error, identical to the previously reported equation for individual aryl sp$^3$ defects.[32]

Figure 5a gives a schematic representation of luminescent oxygen and aryl defects in SWCNTs. Corresponding to the deviation in Raman $\Delta(D/G^+)$ ratio vs $n_d$ slopes, one oxygen defect cluster should correspond to 2-3 individual oxygen defects. This structural model provides a straightforward tool to assess already developed, and in particular, new functionalization methods. If the correlation of Raman $\Delta(D/G^+)$ ratios and calculated defect densities results in values smaller than 172 µm$^{-1}$ (compare Equation (1)), stronger clustering should be expected compared to oxygen defects. Likewise, larger values are an indicator for wider separation of individual structural defects.

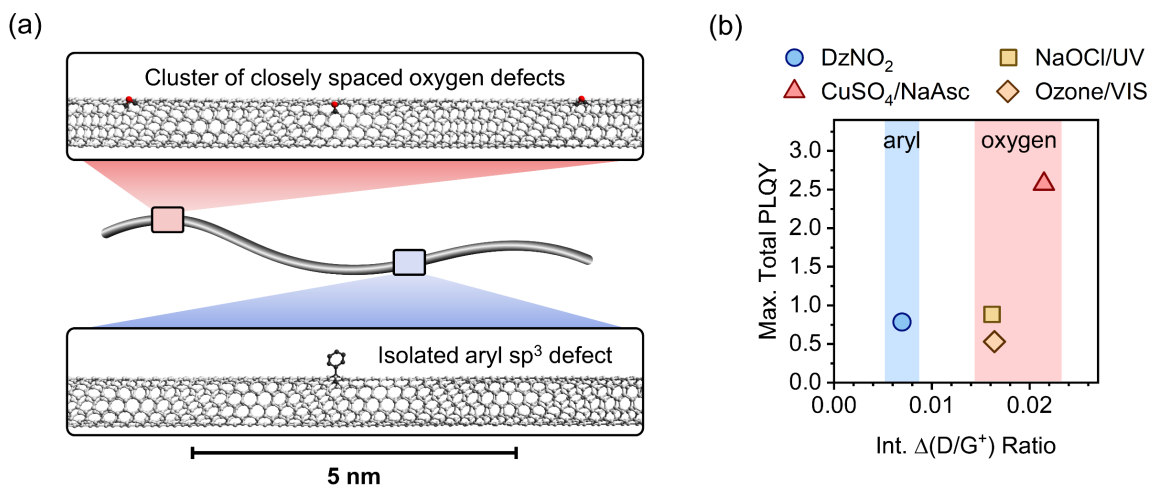

**Figure 5**. **(a)** Schematic depiction of defect clustering in SWCNTs. Closely spaced (smaller than the size of a mobile exciton) luminescent oxygen defects act as one effective exciton trap compared to individual aryl sp$^3$ defects while contributing more to the Raman D/G$^+$ ratio. **(b)** Maximum total PLQYs versus integrated Δ(D/G$^+$) ratio for different functionalization methods. For all luminescent oxygen defects, the highest PLQYs are obtained for larger D/G$^+$ ratios compared to aryl sp$^3$ defects.

For the application of functionalized SWCNTs as emitters the question arises whether defect clustering is desirable or detrimental for reaching the highest total PLQY. Hence, we correlated the maximum total PLQYs for the different functionalization methods with the corresponding Raman Δ(D/G$^+$) ratios (see Figure 5b). Interestingly, the maximum PLQY for SWCNTs with aryl sp³ defects occurs at significantly lower degrees of functionalization compared to oxygen defects. A higher degree of functionalization is required to obtain the brightest SWCNTs with oxygen defects. This might be attributed to the higher actual number of single oxygen defects necessary to achieve the optimal exciton trapping efficiency. The number of oxygen clusters at which maximum total PLQYs are observed are in good agreement with the density of luminescent exciton traps necessary to achieve optimal brightening for aryl-functionalized SWCNTs (~ 4-8 defects µm$^{-1}$). Interestingly, the maximum PLQY (2.5 %) for SWCNTs functionalized with the Fenton-like reaction is substantially higher than those for all other methods (0.5 – 0.9 %). This surprising difference, especially compared to the other oxygen functionalization methods, might be rationalized with the continuous production of low concentrations of ROS by the Fenton-like reaction (with $CuSO_4$/NaAsc) and thus a more controlled reaction with the nanotube lattice. In contrast to that, the harsh conditions of the other methods (*e.g.*, treatment with ozone, hypochlorite or other inorganic oxidizing agents) are more likely to create higher concentrations of ROS and thus may lead to over-functionalized areas that act only as quenching sites and may also cause the observed broadening of the defect emission peaks.[6, 54]

**Conclusion**

In summary, we have shown that the combination of Raman spectroscopy, spectrally-resolved PLQY measurements, and low-temperature single-nanotube PL spectroscopy can be used to recognize and quantify clustering of luminescent oxygen defects in functionalized (6,5) SWCNTs in comparison to typical aryl $sp^3$ defects. One oxygen defect cluster contains about 2-3 individual defects that act as one exciton trap. Since clustering of oxygen defects occurs irrespective of the method of functionalization, we can provide simple and general equations for the precise quantification of oxygen defects and oxygen defect clusters in (6,5) SWCNTs based on standard Raman spectroscopy alone. We show that the comparison of the slopes of the Raman $\Delta(D/G^+)$ ratios vs calculated defect densities from PLQY measurements can serve as a direct measure for clustering. This quantification method can be easily expanded to other functionalization protocols and gives indirect information about the local distribution of exciton trapping sites and individual defects. Hence, this comparative study provides a blueprint for the characterization of new synthetic methods for the introduction of luminescent defects. For example, the functionalization of SWCNTs with bidentate reactants such as bisdiazonium compounds[25] or divalent functional groups[55] could be investigated in more detail based on this approach. A comparison with the reference slopes presented here should enable a quick and reliable assessment of possible defect clustering. Overall, the absolute quantification of defect densities by combining Raman spectroscopy and PLQY measurements is a valuable tool for the systematic and reproducible investigation of luminescent defects in SWCNTs.

**Acknowledgments**

This project has received funding from the European Research Council (ERC) under the European Union's Horizon 2020 research and innovation programme (Grant Agreement No. 817494 "TRIFECTS"). B.S.F. and H.L. gratefully acknowledge support by the DFG under grant numbers FL 834/5-1, FL 834/9-1 and FL 834/12-1. H.L. acknowledges financial support from the Turku Collegium for Science, Medicine and Technology (TCSMT). The authors thank Kerstin Brödner and Marcus Dodds for their technical support during initial tests of the ozone functionalization reaction.

**Data availability**

The data generated and analyzed for this article are available in the heiDATA repository at [link to the repository will be provided prior to publication].

**Author contributions**

F.L.S. fabricated and measured all samples and analyzed the data. S.S. contributed to sample preparation and characterization. H.L. and B.S.F. provided ATPE-sorted (6,5) SWCNTs. J.Z. conceived and supervised the project. F.L.S. and J.Z. wrote the manuscript with input from all authors.

**Competing interests**

The authors declare no competing interests.

# Electronic Supplementary Information (ESI)

# How to Recognize Clustering of Luminescent Defects in Single-Wall Carbon Nanotubes

Finn L. Sebastian[1], Simon Settele[1], Han Li[2,3], Benjamin S. Flavel[4], and Jana Zaumseil[1,*]

[1]Institute for Physical Chemistry, Universität Heidelberg, D-69120 Heidelberg, Germany

[2]Department of Mechanical and Materials Engineering, University of Turku, FI-20014 Turku, Finland

[3]Turku Collegium for Science, Medicine and Technology, University of Turku, FI-20520 Turku, Finland

[4]Institute of Nanotechnology, Karlsruhe Institute of Technology, D-76131 Karlsruhe, Germany

**Corresponding Author**

*E-mail: zaumseil@uni-heidelberg.de

# Experimental Methods

**Dispersion and selection of (6,5) SWCNTs**

(6,5) SWCNTs were selected from CoMoCAT raw material (CHASM SG65i-L58) by aqueous two-phase extraction (ATPE) as reported previously.[1] The raw material was dispersed in deoxycholate (DOC, BioXtra) and transferred to a two-phase system of dextran ($M_W$ = 70 kDa, TCI Chemicals) and poly(ethylene glycol) (PEG, $M_W$ = 6 kDa, Alfa Aesar). For separation of (6,5) SWCNTs, a diameter sorting protocol based on addition of sodium dodecyl sulfate (SDS, Sigma-Aldrich) was employed. The concentration of SDS was increased to 1.1% (w/v) at a fixed DOC concentration of 0.04% (w/v) for removal of large diameter SWCNTs from the top phase. (6,5) SWCNT-enriched fractions were collected between SDS concentration of 1.2% to 1.5% (w/v). Semiconducting and metallic SWCNTs were separated by addition of sodium cholate (SC, Sigma-Aldrich) and sodium hypochlorite (NaOCl, Sigma-Aldrich) as an oxidant. The selected (6,5) SWCNTs in DOC were concentrated by filtration in a pressurized ultrafiltration stirred cell (Millipore) with a $M_W$ = 300 kDa cut-off membrane, and 1% (w/v) SDS was added before further use.

**Functionalization of (6,5) SWCNTs with oxygen and 4-nitroaryl defects**

For the functionalization of (6,5) SWCNTs with oxygen defects, three different methods were used. The first method employed a Fenton-like reaction based on the addition of copper(II) sulfate pentahydrate ($CuSO_4(H_2O)_5$, Sigma-Aldrich, ≥99.9% trace-metal basis) and sodium-L-ascorbate (NaAsc, Sigma-Aldrich, ≥99%) according to a procedure by Settele et al.[2] In short, aqueous dispersions of (6,5) SWCNT were adjusted to an optical density (OD) of 0.33 cm$^{-1}$ at the $E_{11}$ absorption peak with ultra-pure water (total volume 1 ml), and a final surfactant concentration of 0.33% (w/v) SDS. Aliquots of aqueous solutions of $CuSO_4$ and NaAsc were added to (6,5) SWCNT dispersions to achieve final concentrations of $CuSO_4$ between 25 and 250 µM and a molar ratio of 1:12 ($CuSO_4$:NaAsc) in the reaction mixture. After storage in the dark for 16 h, the reaction was stopped by addition of $Na_4EDTA$ (10 µl, 1.4 M, Sigma-Aldrich, 98%) and 10% (w/v) DOC (20 µl), followed by irradiation with UV-light for 2 h (365 nm, SOLIS-365C, Thorlabs, 1.9 mW·mm$^{-2}$) to promote the reorganization of oxygen defects.

The second functionalization method was adapted from a protocol reported by Lin et al.[3] ATPE-sorted dispersions of (6,5) SWCNTs in 1% (w/v) SDS were diluted to an OD of 0.1 cm$^{-1}$ (total volume 3 ml) and a final SDS concentration of 0.1% (w/v). NaOCl (12% available chloride, Carl Roth) was added to achieve final concentrations between 0.015 and 1 mM in the reaction mixture. After irradiation with UV-C-light (254 nm, VL-215.LC, Vilber Lourmat, 15 W) for 30 min, the reaction was stopped by addition of 20 µl of 10% (w/v) aqueous DOC solution.

The third functionalization method was adapted from Ghosh et al.[4] Ozone (Ozone Generator, Cambridge NanoTech Inc.) was bubbled through 7 ml of ultra-pure water in a Schlenk tube for 10 min, followed by dilution to an OD of 0.3 cm$^{-1}$ at the ozone UV absorption peak (260 nm). ATPE-sorted dispersions of (6,5) SWCNTs were adjusted to an OD of 0.2 cm$^{-1}$ (total volume 1.5 ml) and a final SDS concentration of 0.2 cm$^{-1}$. Depending on the intended degree of functionalization, between 5 and 50 µl of the ozonated ultra-pure water were slowly added while stirring. Photoconversion of the reaction product was achieved by irradiation with a green light-emitting diode (525 nm, SOLIS-525C, Thorlabs, 1.5 mW·mm$^{-2}$) until no further changes of the PL spectrum were observed (~ 60-90 min).

For functionalization of SWCNTs with 4-nitroaryl defects, ATPE-sorted (6,5) SWCNTs were diluted to an OD of 0.33 cm$^{-1}$ at the $E_{11}$ absorption peak (total volume 1 ml) and an SDS concentration of 0.33 cm$^{-1}$. Aliquots of aqueous solutions of 4-nitrobenzenediazonium tetrafluoroborate (DzNO$_2$, TCI Chemicals, >98%) were added to achieve concentrations between 0.04 and 0.42 µM in the reaction mixture. After storage in the dark for 7 days to ensure complete decomposition of the diazonium salt, 20 µl of 10% (w/v) aqueous DOC solution were added to the reaction mixture.

Independent of the functionalization method, SWCNT dispersions were spin-filtrated (Amicon Ultra-4, 100 kDa cut-off) immediately after completion of the reaction and transferred to 1% (w/v) aqueous DOC before further characterization.

**Chemical doping of (6,5) SWCNTs films**

Drop-cast films of (6,5) SWCNTs on glass (Schott AF32eco) were immersed in solutions of the molecular p-type dopant 2,3,5,6-tetrafluoro-7,7,8,8-tetracyanoquinodimethane (F$_4$TCNQ, TCI Chemicals, >98%) in toluene (200 µg·ml$^{-1}$) for different periods of time. The films were rinsed with toluene and blow-dried with nitrogen before further characterization.

**Transfer of aqueous (6,5) SWCNT dispersions to toluene**

A volume of 5 ml of functionalized (6,5) SWCNTs in 0.33% (w/v) aqueous SDS solution were filtered through PTFE membranes (Merck Millipore JVWP, pore size 0.1 µm) and washed thoroughly with ultra-pure water to remove the surfactant. The filter was then subjected to bath sonication for 30 min in a solution of poly[(9,9-dioctylfluorenyl-2,7-diyl)-*alt*-(6,6′-(2,2′-bipyridine))] (PFO-BPy, American Dye Source, $M_W$ = 40 kDa, 0.5 g·L$^{-1}$) in toluene (5 ml, Sigma-Aldrich, ≥99.5%). Unexfoliated material was removed by centrifugation of the stock dispersion, followed by centrifugation of the supernatant (each step at 60000 $g$ for 45 min, Beckman Coulter Avanti J26XP centrifuge). The supernatant obtained after the second step was used to prepare samples for AFM and low-temperature PL spectroscopy.

**Characterization**

Baseline-corrected UV-Vis-nIR absorption spectra were recorded with a Cary 6000i absorption spectrometer (Varian, Inc.). A scattering background $S(\lambda) = S_0 e^{-b\lambda}$ was subtracted from the acquired spectra.[5]

Atomic force microscopy (AFM) was performed under ambient conditions with a Bruker Dimension Icon atomic force microscope in ScanAsyst mode. The SWCNTs were spin-coated (2000 rpm, 60 s) onto cleaned native silicon wafers from dispersions obtained after transfer of SDS-dispersed SWCNTs to toluene/PFO-BPy, followed by rinsing with tetrahydrofuran (THF) and isopropanol (IPA). Gwyddion 2.64 was used for nanotube length distribution analysis.

Raman spectra of pristine and functionalized (6,5) SWCNTs were acquired with a Renishaw InVia Reflex confocal Raman microscope in backscattering configuration (50× long working distance objective, Olympus, N.A. 0.5) under near-resonant excitation with a 532 nm laser. Dispersions of SWCNTs were drop-cast onto glass substrates (Schott AF32eco), and gently rinsed with water. Over 3600 spectra were collected and averaged for each sample. Integrated Raman $\Delta(D/G^+)$ ratios were determined by integration of the average normalized Raman spectra from 1200 – 1400 cm$^{-1}$ (D-mode) and 1560 – 1650 cm$^{-1}$ ($G^+$-mode).

Room-temperature photoluminescence (PL) spectroscopy and emission-excitation (PLE) mapping was performed with a Fluorolog-3 spectrometer (Horiba Jobin-Yvon). A xenon arc-discharge lamp (450 W) coupled to a double grating monochromator was employed as the excitation light source. Light emitted from the SWCNT dispersions was passed through a long-pass filter (850 nm), Czerny-Turner imaging spectrograph and recorded with a liquid nitrogen cooled InGaAs line camera (Symphony II).

**Photoluminescence quantum yields**

Absolute photoluminescence quantum yields (PLQYs) of pristine and functionalized dispersions were determined with an integrating sphere as the ratio of emitted ($N_{em}$) and absorbed photons ($N_{abs}$) in accordance with previously reported procedures:[6, 7]

$$\eta = \frac{N_{em}}{N_{abs}} \qquad (1)$$

SWCNT dispersions were diluted to an OD of 0.15 to 0.2 cm$^{-1}$ at the $E_{11}$ transition to minimize re-absorption. A quartz glass cuvette (Hellma Analytics QX) with 1 ml of dispersion was placed in an integrating sphere (LabSphere, Spectralon coating). The spectrally filtered output of a picosecond-pulsed supercontinuum laser (NKT Photonics SuperK Extreme) was used for resonant excitation of SWCNTs at the $E_{22}$ transition (570 nm), and the light exiting the integrating sphere was coupled into the spectrometer (Acton SpectraPro SP2358) with an optical fiber. The PL of the sample in the near-infrared (nIR) and the attenuated laser signal at the excitation wavelength were recorded with a liquid nitrogen-cooled InGaAs line camera (Princeton Instruments OMA V:1024). Identical measurements were performed with 1% (w/v) aqueous DOC to correct for light absorption and scattering of the solvent. Integration of the sample emission yields a value proportional to the number of emitted photons. Similarly, the integrated attenuated laser signal for each sample is subtracted from the value obtained for the solvent reference to determine the number of absorbed photons. To account for the wavelength-dependent sensitivity of the detector and the absorption of optical components, a stabilized broadband light source (Thorlabs SLS201/M, 300-2600 nm) with known spectral output was used to collect calibration lamp spectra.

**Calculation of luminescent defect density**

Luminescent defect densities were calculated based on the model of diffusion-limited contact quenching (DLCQ) of excitons in SWCNTs as reported previously.[1] The DLCQ model assumes that $E_{11}$ excitons undergo either nonradiative quenching at quenching sites, *e.g.*, nanotube ends, or decay *via* radiative $E_{11}$ PL emission. The PLQY of the $E_{11}$ emission in pristine SWCNT can be calculated as

$$\eta = \frac{\pi}{2 \cdot n_q^2 \cdot D \cdot \tau_{rad}}, \qquad (2)$$

where $n_q$ represents the number of quenching sites per μm of nanotube, $D$ is the exciton diffusion constant and $\tau_{rad}$ the radiative lifetime of the $E_{11}$ exciton. Here, values of $D = 10.7 \pm 0.4$ cm$^2 \cdot$s$^{-1}$ and $\tau_{rad} = 3.35 \pm 0.41$ ns are used as reported in previous experimental studies.[8, 9] The introduction of luminescent defects ($n_d$) leads to additional radiative decay pathways resulting in a reduction of the $E_{11}$ PLQY $\eta^*$ of functionalized SWCNTs:

$$\eta^* = \frac{\pi}{2 \cdot (n_q + n_d)^2 \cdot D \cdot \tau_{rad}} \qquad (3)$$

The combination of equations (2) and (3) yields an expression for the number density (µm$^{-1}$) of luminescent defects:

$$n_d = n_q \left( \sqrt{\frac{\eta}{\eta^*}} - 1 \right) = \sqrt{\frac{\pi}{2 \cdot \eta \cdot D \cdot \tau_{rad}}} \left( \sqrt{\frac{\eta}{\eta^*}} - 1 \right) \qquad (4)$$

**Single-SWCNT PL spectroscopy at cryogenic temperatures**

Dispersions obtained after transfer of SDS-dispersed SWCNTs to toluene/PFO-BPy were diluted to an OD of 0.005 cm$^{-1}$ at the $E_{11}$ absorption peak and mixed with a solution of polystyrene (Polymer Source Inc., $M_W$ = 230 kDa) in toluene (40 g·L$^{-1}$) in equal volumes. Spin-coating (2000 rpm, 60 s) of the obtained mixture (30 µl) onto gold-coated (150 nm) glass substrates completed the sample preparation.

A closed-cycle liquid helium cooled optical cryostat (Montana Instruments, Cryostation s50) was employed to perform low-temperature single-SWCNT PL spectroscopy at 4.7 K. To this end, a nIR-optimized ×50 long-working distance objective (Mitutoyo, N.A. 0.42) was used to focus the output of a continuous wave laser diode (Coherent, Inc. OBIS 640 nm, 1 mW) onto the sample. Emitted light was collected and projected through a long-pass filter (850 nm) onto a grating spectrograph (Princeton Instruments, IsoPlane SCT-320) equipped with a 1200 nm blaze grating (85 lines per mm). PL spectra were recorded with a thermoelectrically cooled two-dimensional InGaAs camera array (Princeton Instruments, NIRvana 640ST). The data collection was automated using a custom script to control a piezo-based nanopositioning system (Attocube ANC350, ANPx101 and ANPz102 nanopositioners). A small number of PL spectra probably originating from bundled or aggregated SWCNTs were identified by the presence of broad and ill-defined $E_{11}$ PL emission in accordance with literature reports.[1,10,11] These spectra were disregarded in the subsequent statistical analysis.

# Additional spectra

**UV-Vis-nIR absorption spectrum of pristine (6,5) SWCNTs**

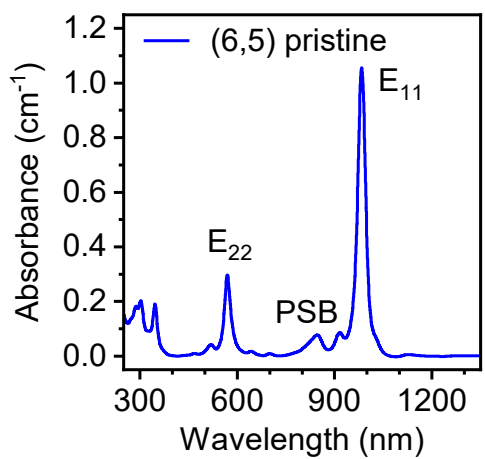

**Figure S1**. Baseline-corrected UV-Vis-nIR absorption spectrum of pristine (6,5) SWCNTs in aqueous dispersion (1% (w/v) SDS), with assignment of the $E_{22}$, $E_{11}$, and $E_{11}$ phonon sideband (PSB) optical transitions.

**Photoluminescence excitation-emission map of pristine (6,5) SWCNTs**

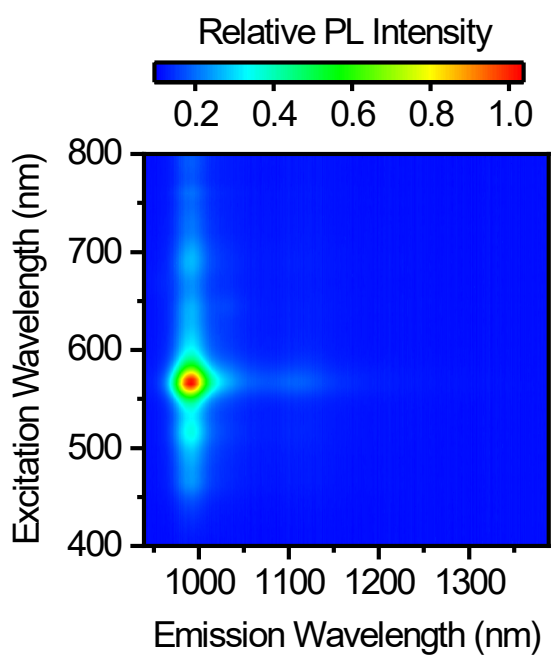

**Figure S2**. Photoluminescence excitation-emission map of pristine (6,5) SWCNTs in aqueous dispersion (0.3% (w/v) SDS).

**UV-Vis absorption spectrum of ozonated water**

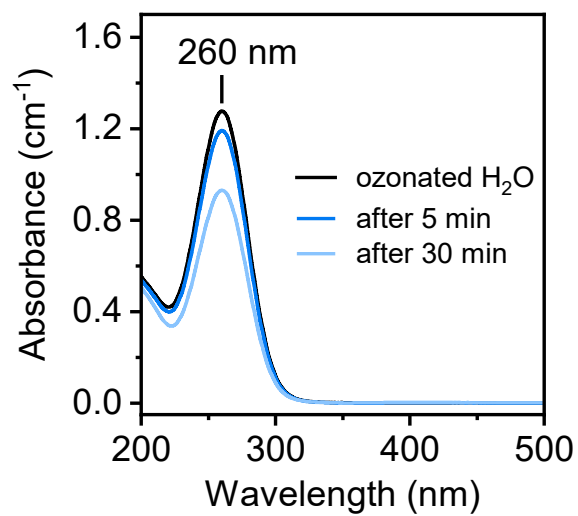

**Figure S3**. UV-Vis-nIR absorption spectra of water after ozonation, demonstrating the evolution of the ozone absorption peak (260 nm) 5 min and 30 min after ozonation.

**Absolute PL spectra for different functionalization methods**

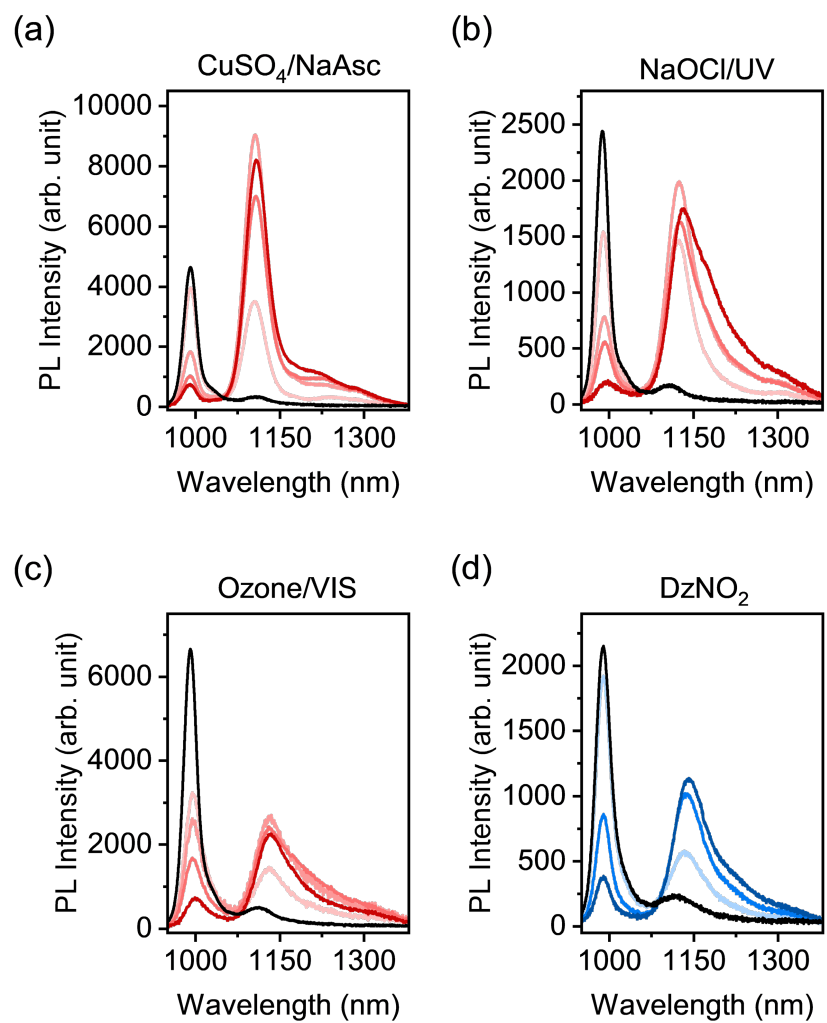

**Figure S4**. Absolute PL spectra of (6,5) SWCNTs functionalized with luminescent oxygen and aryl defects by different synthetic methods (**(a)** CuSO$_4$/NaAsc, **(b)** NaOCl/UV-light, **(c)** ozone/visible light, **(d)** DzNO$_2$).

**E$_{11}$*/E$_{11}$ PL intensity ratios for different functionalization methods**

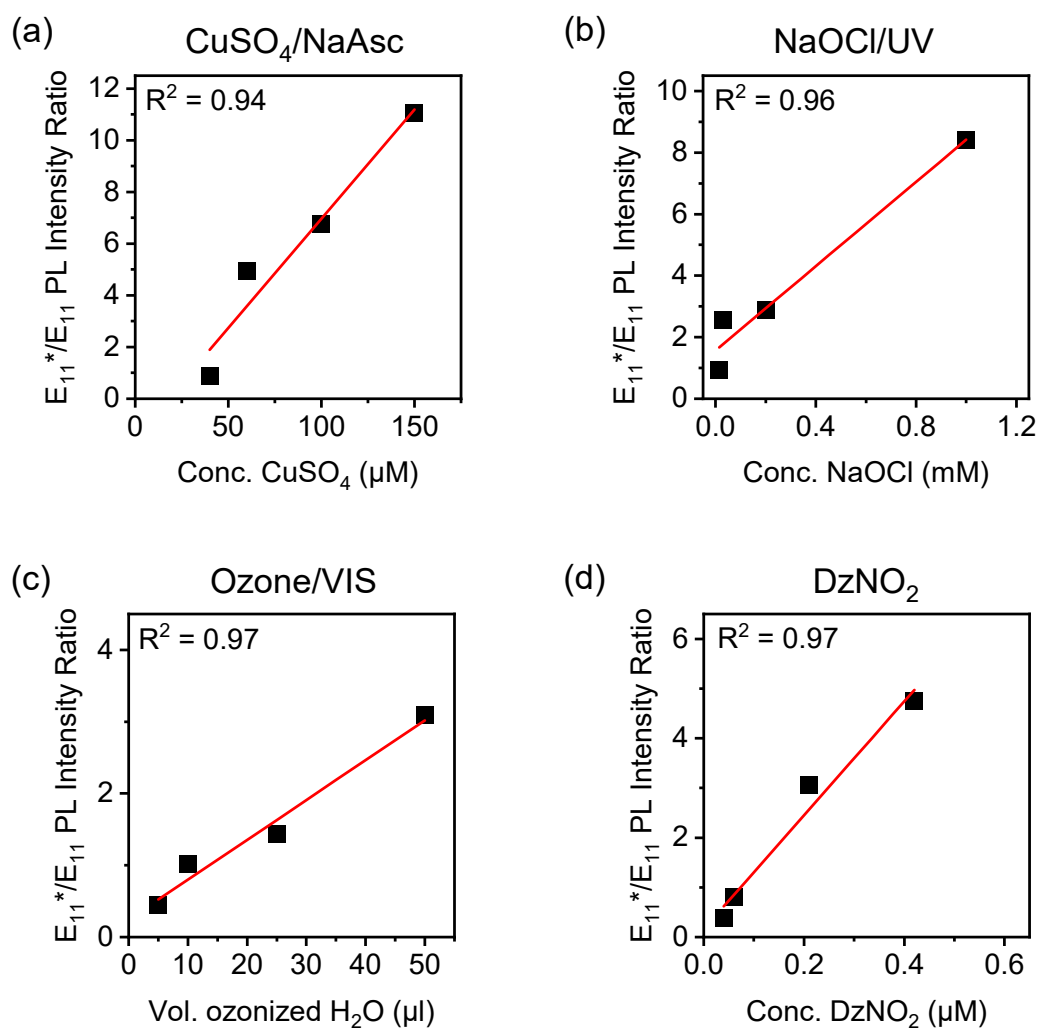

**Figure S5**. Dependence of the E$_{11}$*/E$_{11}$ PL intensity ratios on the reactant concentration for (6,5) SWCNTs functionalized with luminescent oxygen and aryl defects by different synthetic methods (**(a)** CuSO$_4$/NaAsc, **(b)** NaOCl/UV-light, **(c)** ozone/visible light, **(d)** DzNO$_2$). Red lines are linear fits to the data (respective $R^2$ values are noted in the graphs).

**Raman spectra of p-doped pristine and functionalized SWCNTs**

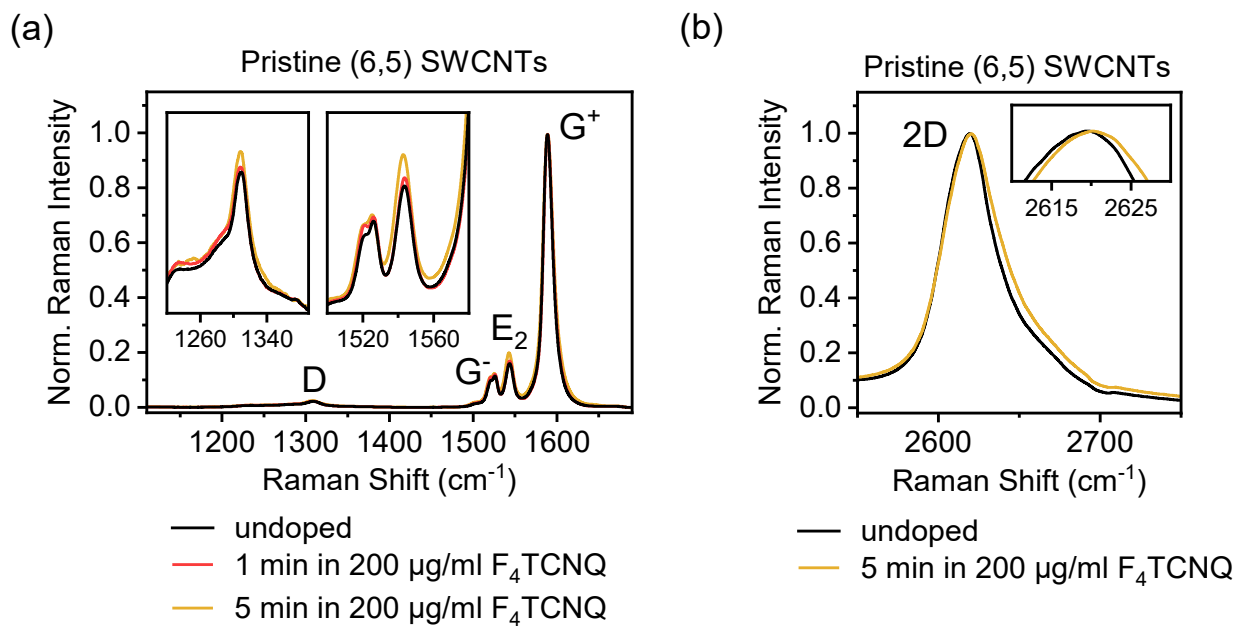

**Figure S6**. Influence of chemical doping on Raman spectra of (6,5) SWCNTs. **(a)** Averaged and normalized (to $G^+$-mode) Raman spectra of pristine (6,5) SWCNTs (drop cast film) without treatment and after 1 min and 5 min of immersion in a solution of the p-type dopant F4TCNQ (solution in toluene, 200 µg/ml). **(b)** Averaged and normalized Raman spectra of the 2D-mode of pristine (6,5) SWCNTs before and after doping.

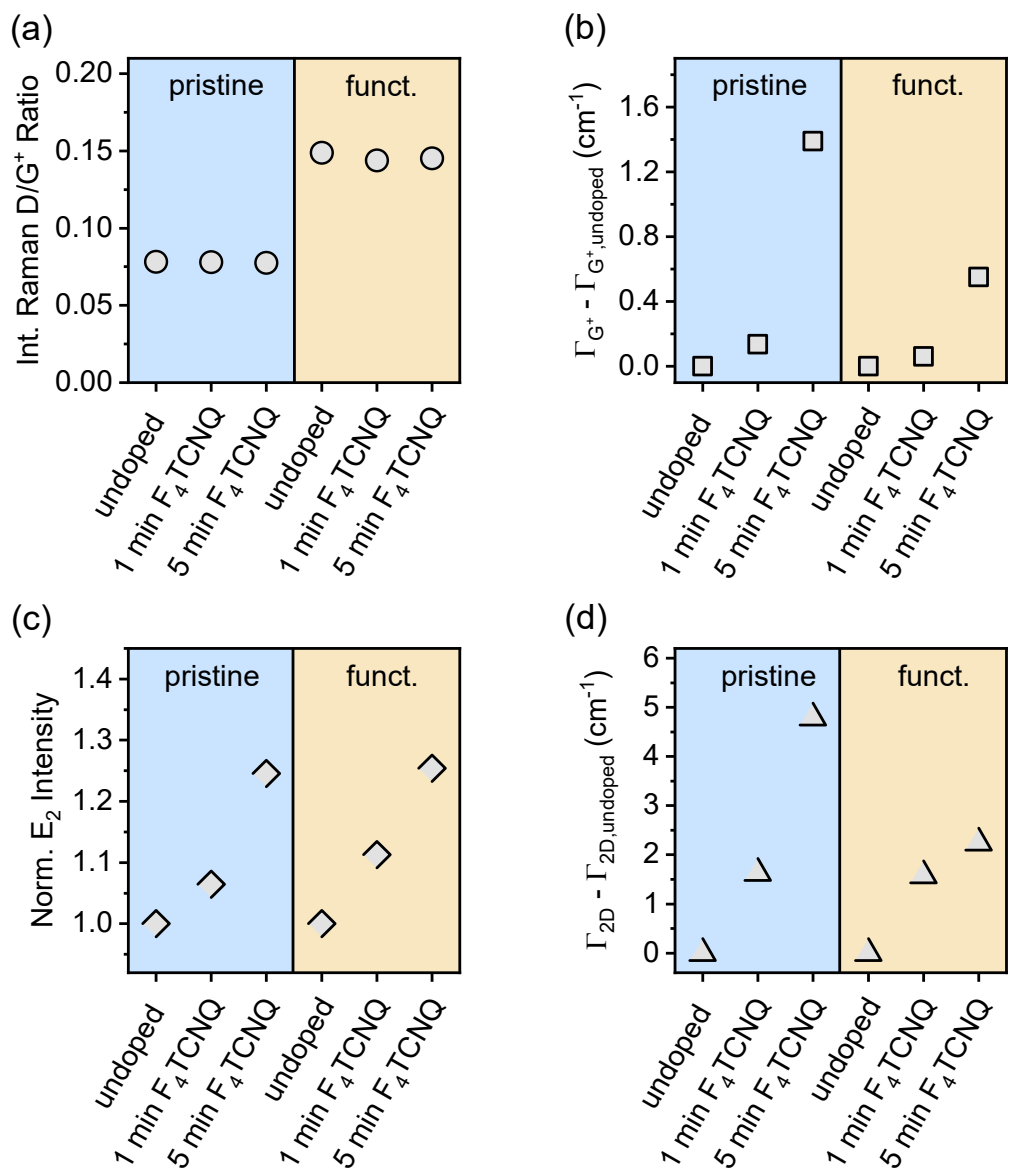

**Figure S7**. **(a)** Integrated Raman D/G$^+$ ratios for undoped/doped SWCNTs (pristine and functionalized with CuSO$_4$/NaAsc). **(b)** Raman G$^+$-mode linewidth ($\Gamma_{D+}$, full width at half maximum) for doped SWCNTs relative to the initial linewidth of the undoped sample. **(c)** Normalized intensity of the Raman E$_2$-mode for undoped/doped SWCNTs. **(d)** Linewidth of Raman 2D-mode ($\Gamma_{2D}$) for doped SWCNTs relative to the initial linewidth of the undoped sample.

**Raman Δ(D/G⁺) *vs* calculated defect density *n*<sub>d</sub> for different functionalization methods**

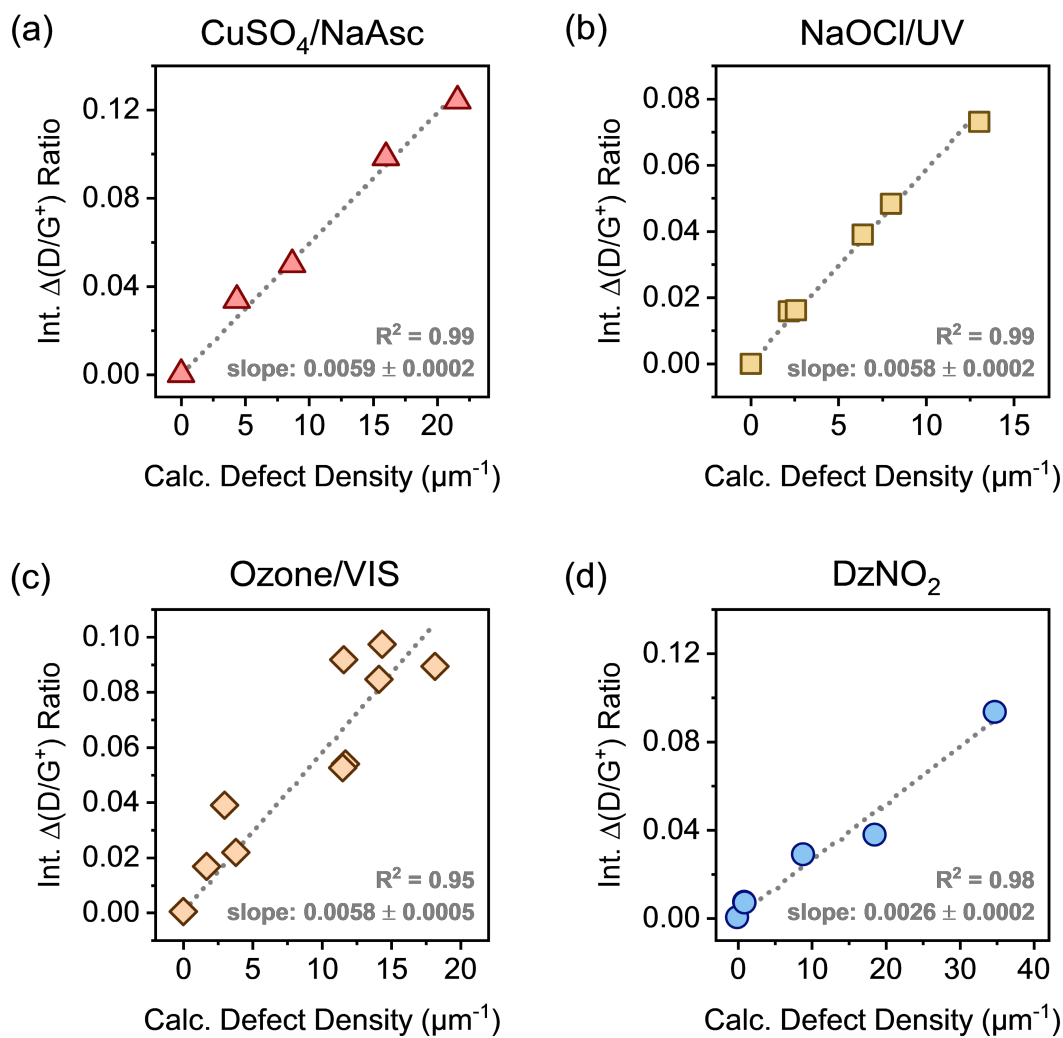

**Figure S8**. Integrated Raman $\Delta(D/G^+)$ ratios versus calculated defect density $n_d$ ($\mu m^{-1}$) from PLQY measurements for (6,5) SWCNTs functionalized with luminescent oxygen and aryl defects via different synthetic methods (**(a)** CuSO$_4$/NaAsc, **(b)** NaOCl/UV-light, **(c)** ozone/visible light, **(d)** DzNO$_2$). Dashed grey lines are linear fits to the data (slope of the fit and respective $R^2$ values are noted in the graphs).

**Raman spectroscopy of pristine and functionalized SWCNTs (RBM and IFM region)**

The impact of luminescent oxygen and aryl defects on Raman modes other than the D- and $G^+$-mode should also be evaluated. In the spectral region between the radial breathing mode (RBM) and D-mode of SWCNTs, so-called intermediate frequency modes (IFMs) with weaker absolute signal intensities can be identified. For some IFMs dispersive characteristics in accordance with the involvement of double-resonant scattering processes were reported that are indicative of a defect-related origin.[12] Recently, we have demonstrated that the intensities of those IFMs, which can be derived from the ZA and ZO phonon branches in graphene, increase linearly with the defect density calculated from the $E_{11}$ PLQY. This relation provides an additional method for the absolute quantification of luminescent defects in SWCNTs.[13] Furthermore, due to their chirality-dependent frequency, defect-related IFMs can also be applied for a relative quantification of the defect density in SWCNT mixtures.[14]

We recorded Raman spectra of drop-cast films of oxygen- and aryl-functionalized (6,5) SWCNTs in the IFM region using a 785 nm excitation laser. Averaged Raman spectra for all functionalization methods are shown in **Figure S9**. In all spectra the ZA-derived IFM can be identified at ~480 $cm^{-1}$ whereas the ZO-derived IFM appears at ~580 $cm^{-1}$. For a quantitative evaluation the integrated IFM intensities were divided by the integrated intensity of the RBM of (6,5) SWCNTs (308 $cm^{-1}$) and the change of the integrated ratio, Δ(IFM/RBM), was correlated with the calculated defect density $n_d$. Irrespective of the functionalization method, a linear relation between Δ(IFM/RBM) and $n_d$ was found although with significantly steeper slopes for oxygen defects compared to aryl defects, similar to the results reported for the $D/G^+$ ratios (Figure 3a). Individual slopes of Δ(IFM/RBM) *vs* $n_d$ are presented in **Figure S10**. The low absolute Raman intensities of IFMs lead to some deviations of the Δ(IFM/RBM) *vs* $n_d$ slopes for different methods of oxygen defect functionalization. However, the overall trend clearly supports the findings for the Raman Δ($D/G^+$) ratio. For both the slope of Δ(ZA/RBM) and Δ(ZO/RBM) *vs* $n_d$, the values are larger by a factor of 1.9 for oxygen defects compared to aryl defects (see **Figure S11**).

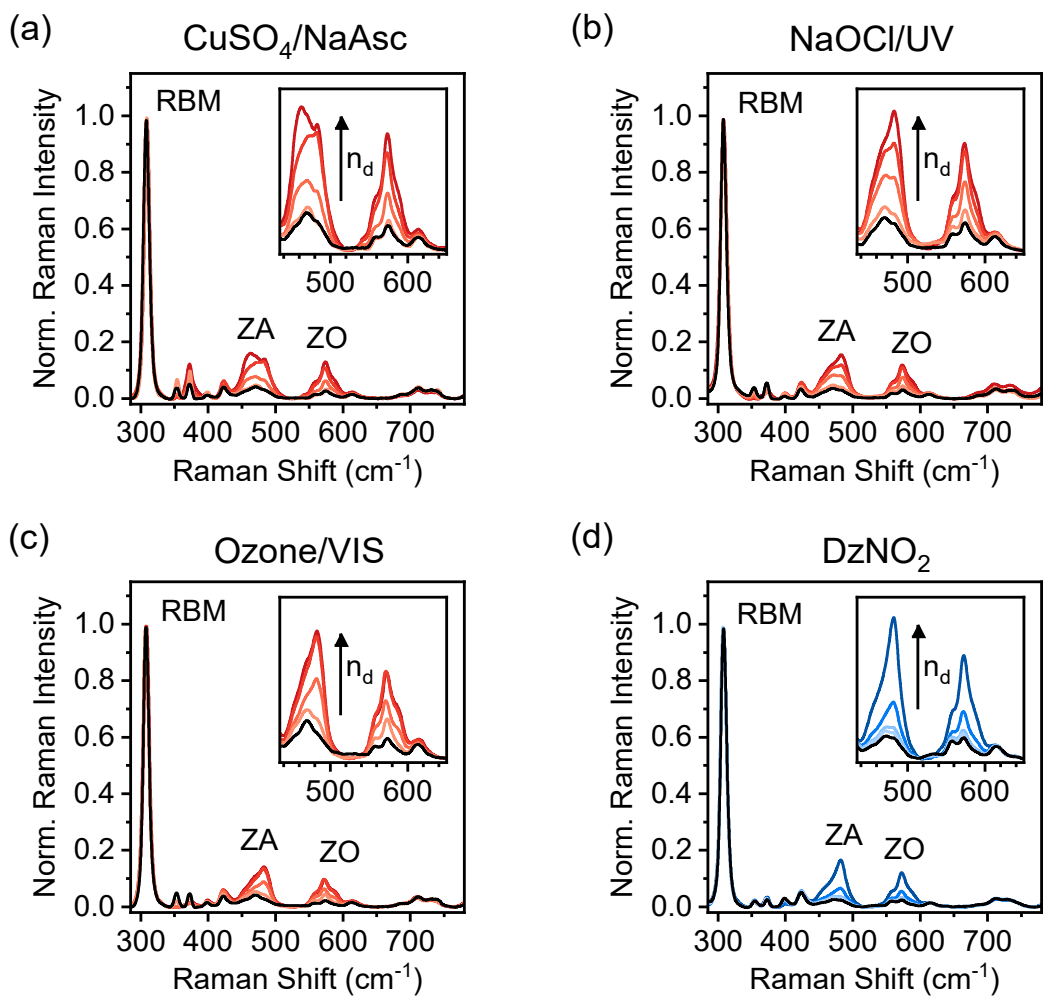

**Figure S9.** Averaged and normalized (to RBM at 308 cm$^{-1}$) Raman spectra of pristine and functionalized (6,5) SWCNTs for all functionalization methods; **(a)** CuSO$_4$/NaAsc, **(b)** NaOCl/UV-light, **(c)** ozone/visible light, **(d)** DzNO$_2$, $\lambda_{exc}$ = 785 nm, >3600 individual spectra per sample. Insets display the evolution of the intermediate frequency modes (IFMs) for increasing degree of functionalization.

**Raman Δ(IFM/RBM) *vs* calculated defect density $n_d$ for different functionalization methods**

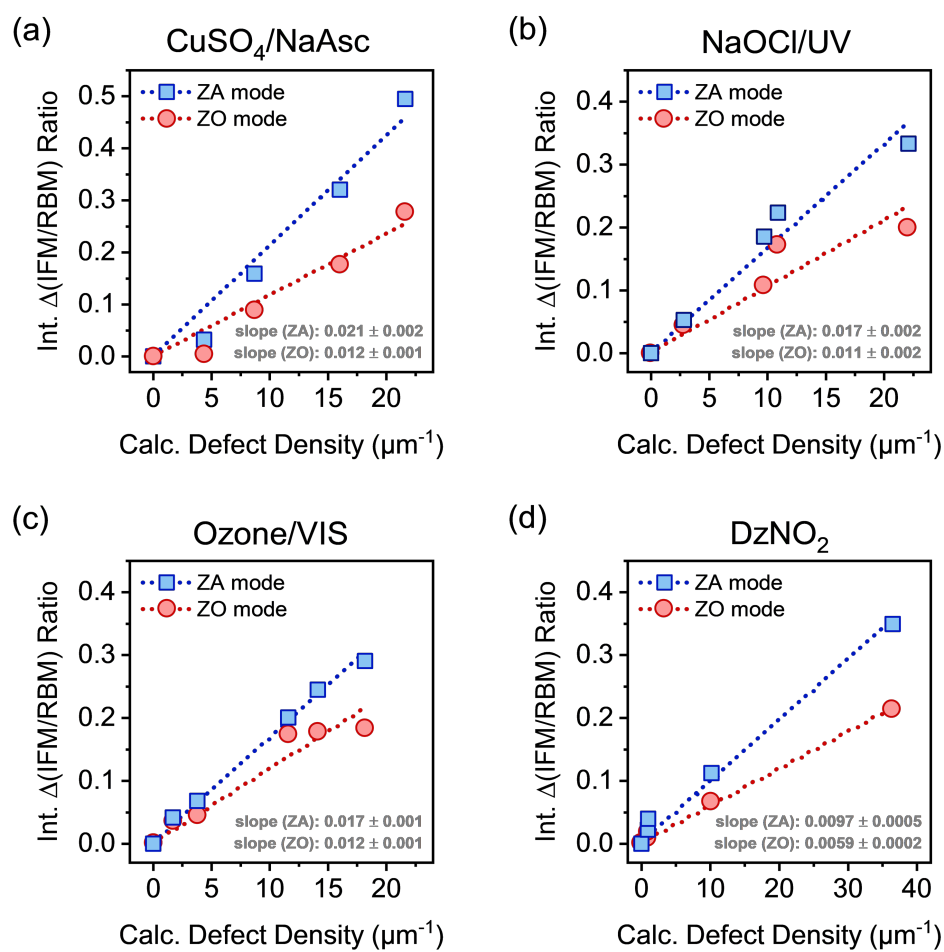

**Figure S10**. Integrated Raman Δ(ZA/RBM) and Δ(ZO/RBM) ratios versus calculated defect densities $n_d$ for pristine and functionalized (6,5) SWCNTs (drop cast films) for all functionalization methods; **(a)** CuSO$_4$/NaAsc, **(b)** NaOCl/UV-light, **(c)** ozone/visible light, **(d)** DzNO$_2$. Dashed lines are linear fits to the data (slopes of the fits and respective $R^2$ values are noted in the graphs).

**Extracted Raman Δ(IFM/RBM) *vs* n_d slopes**

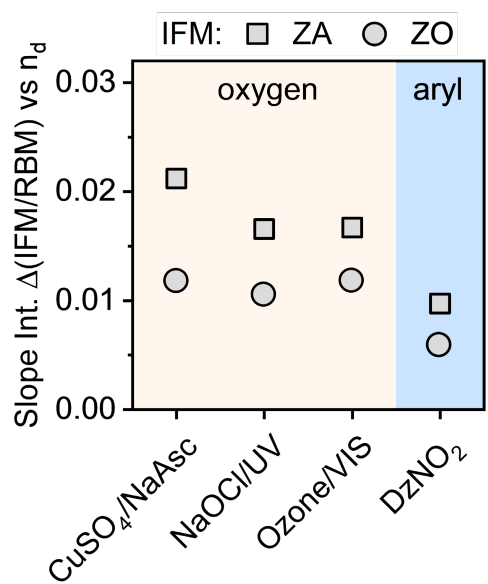

**Figure S11.** Extracted slopes of the integrated Δ(IFM/RBM) ratios versus calculated defect densities $n_d$ for different functionalization methods.

## Additional low-temperature single-SWCNT PL spectra

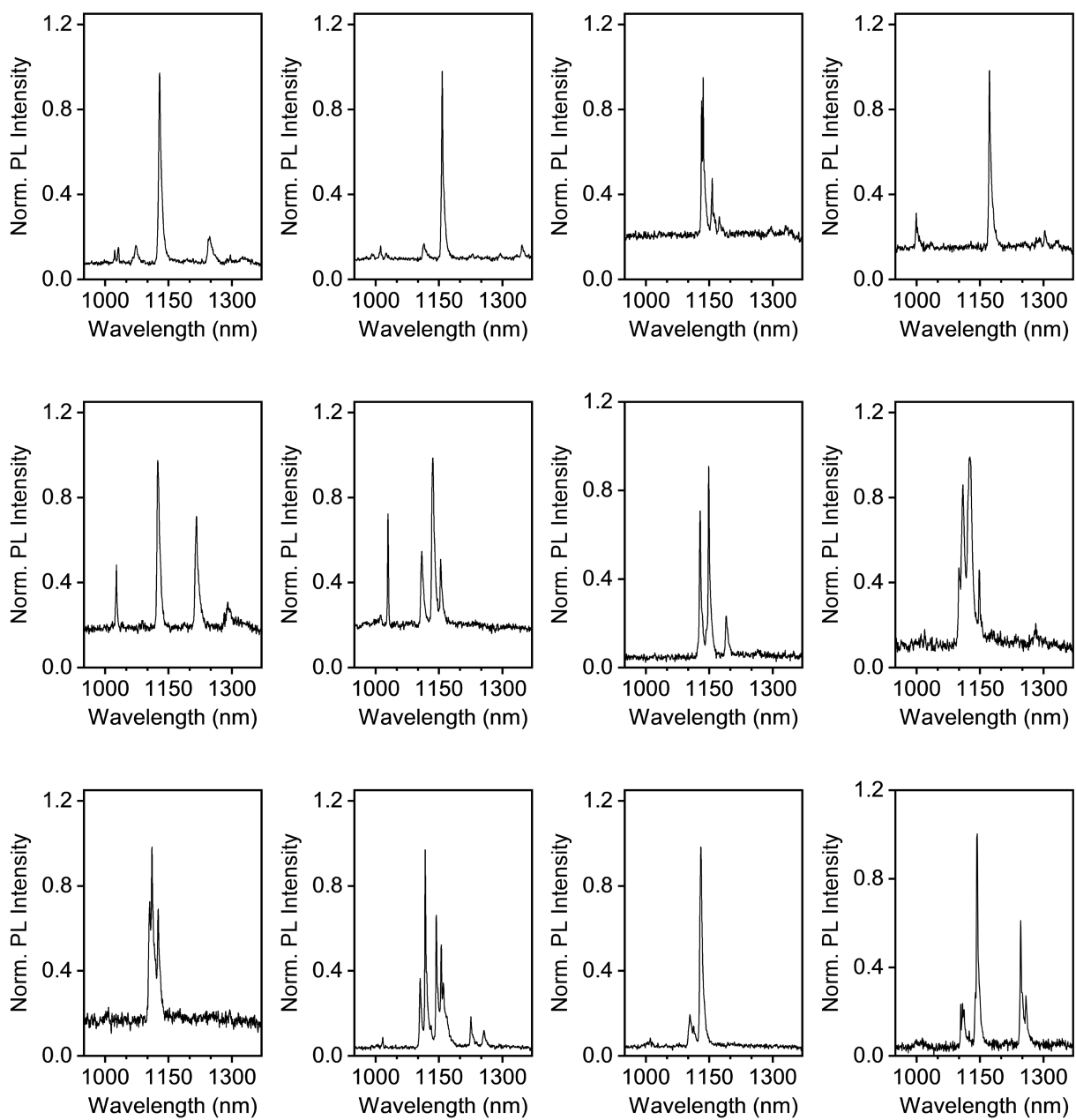

**Figure S12**. Additional low-temperature (4.7 K) PL spectra of single SWCNTs functionalized with $CuSO_4$/NaAsc, embedded in a polystyrene matrix (continued on the next page).

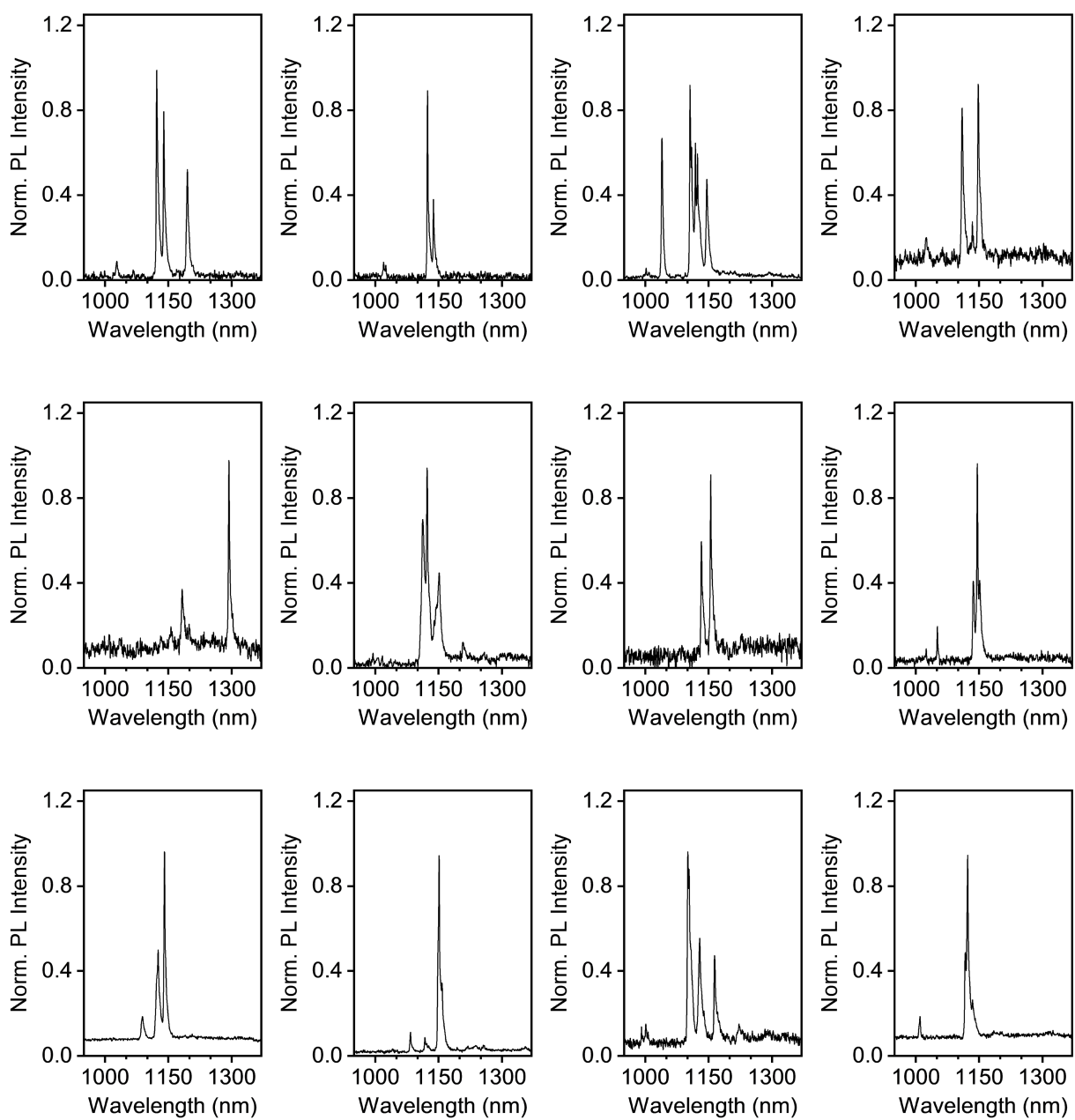

**Figure S13**. Additional low-temperature (4.7 K) PL spectra of single SWCNTs functionalized with CuSO$_4$/NaAsc, embedded in a polystyrene matrix.

**AFM length statistics of functionalized SWCNTs**

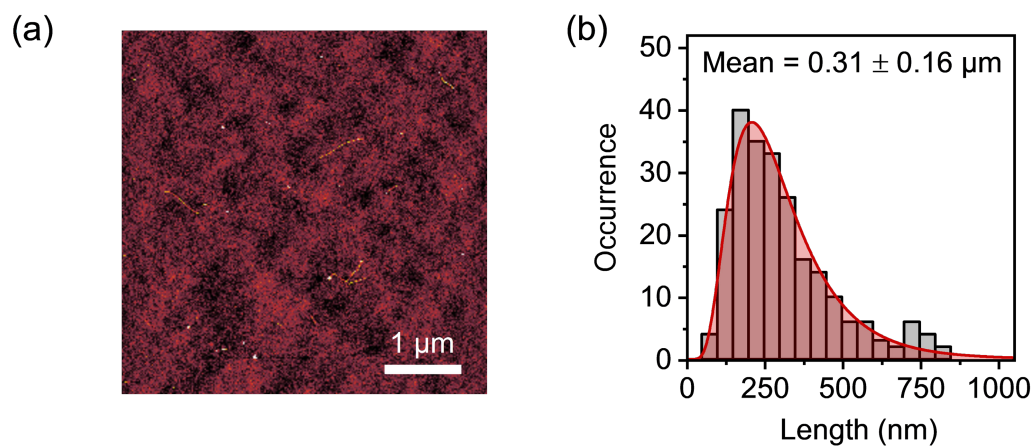

**Figure S14.** **(a)** Representative Atomic force micrograph and **(b)** length histogram (log-normal distribution, red line) of functionalized (6,5) SWCNTs (CuSO$_4$/NaAsc) after transfer from aqueous dispersion (0.3 % w/v SDS) to toluene/PFO-BPy.